\documentclass[12pt,preprint]{aastex}

\shortauthors{Savin et al.} 
\shorttitle{Rate Coefficient for H$^+$ + H$_2$ Charge Transfer and Some
Cosmological Implications}

\begin{document}

\title{Rate Coefficient for {\boldmath ${\rm H}^+ + {\rm
H}_2(X\ ^1\Sigma_g^+,\ \nu=0,\ J=0)  \to {\rm H}(1s) + {\rm H}_2^+$} Charge
Transfer and Some Cosmological Implications}

\author{
Daniel Wolf Savin\altaffilmark{1},
Predrag S.\ Krsti\'c\altaffilmark{2},
Zolt\'an Haiman\altaffilmark{3}, and
Phillip C.\ Stancil\altaffilmark{4}
}

\altaffiltext{1}{Columbia Astrophysics Laboratory, Columbia University, 
New York, NY 10027-6601; savin@astro.columbia.edu}
\altaffiltext{2}{Physics Division, Oak Ridge National Laboratory,
Oak Ridge, TN 37831-6372; krsticp@ornl.gov}
\altaffiltext{3}{Department of Astronomy, Columbia University, New York, 
NY 10027-6601; zoltan@astro.columbia.edu}
\altaffiltext{4}{Department of Physics and Astronomy and Center for 
Simulational Physics, University of Georgia, Athens, GA 30602-2452; 
stancil@physast.uga.edu}

\begin{abstract} 

Krsti\'c has carried out the first quantum mechanical calculations
near threshold for the charge transfer (CT) process ${\rm H}^+ + {\rm
H}_2(X\ ^1\Sigma_g^+,\ \nu=0,\ J=0) \to {\rm H}(1s) + {\rm H}_2^+$.  These
results are relevant for models of primordial galaxy and first star
formation that require reliable atomic and molecular data for
obtaining the early universe hydrogen chemistry.  Using the results of
Krsti\'c, we calculate the relevant CT rate coefficient for temperatures
between 100 and 30,000~K.  We also present a simple fit which can be
readily implemented into early universe chemical models.  Additionally, we
explore how the range of previously published data for this reaction
translates into uncertainties in the predicted gas temperature and H$_2$
relative abundance in a collapsing primordial gas cloud.  Our new data
significantly reduce these cosmological uncertainties that are due to the
uncertainties in the previously published CT rate coefficients.

\end{abstract}

\keywords{atomic data --- early universe --- galaxies: formation ---
molecular data --- molecular processes --- stars: formation}

\section{Introduction}
\label{sec:Intro}

Molecular hydrogen is an important coolant during the epoch of
primordial galaxy and first star formation.  Model calculations
indicate that radiatively-induced cooling due to collisions of H$_2$,
primarily with H, dominate the cooling of collapsing primordial gas
clouds from temperatures beginning at $\sim 10^4$~K and going down to
$\sim 5\times 10^2$~K \citep[e.g.,][]{Sas67a,Abe00a,Nak02a}.

The importance of H$_2$ in the early universe is supported by the
recent {\it Wilkinson Microwave Anisotropy Probe} ({\it WMAP})
measurements which suggest (a) that reionization occured at high
redshift and (b) a large Thompson scattering optical depth.  These
results require star formation at this redshift to produce the
ionizing radiation, which implies that H$_2$ cooling is important for
early star formation \citep{Spe03a,Hai03a}.  Given the pivotal role
that H$_2$ is predicted to play, an accurate understanding of its
formation and destruction in the early universe is crucial for
understanding the formation of hierarchical structure (Haiman, Thoul,
\& Loeb 1996; Abel et al.\ 1997; Abel, Bryan, \& Norman 2002; Galli \&
Palla 1998; Flower 2002; Lepp, Stancil, \& Dalgarno 2002; Nakamura \&
Umemura 2002; Ricotti, Gnedin, \& Shull 2002).

The early universe chemistry of H$_2$ during the epoch of primordial
galaxy and first star formation has been reviewed recently by a number
of groups \citep[e.g.,][]{Abe97a,Gal98a,Lep02a,Oh02a}.  These authors
have found that the relative H$_2$ abundances are determined by only a
handful of processes.  Of particular interest, \citet{Oh02a} point out
that the charge transfer (CT) process
\begin{equation}
\label{eq:CTH2}
{\rm H}^+ + {\rm H}_2(X\ ^1\Sigma_g^+,\ \nu=0,\ J=0)  \to {\rm H}(1s) + {\rm 
H_2}^+
\end{equation}
is the dominant destruction mechanism of H$_2$ during the formation at
this epoch of halos with viral temperatures $T_{vir}>10^4$~K.  This
process dominates until a collapsing cloud of primordial gas begins
internally to emit ionizing radiation or until the background
extragalactic UV radiation field has reached significant levels.  Then
the dominant destruction mechanism becomes photodissociation through
the Lyman-Werver bands via the two-step Solomon process (Haiman, Rees,
\& Loeb 1997).

Because of the fundamental importance of reaction~(\ref{eq:CTH2}),
reliable rate coefficients for this process are needed for our
understanding of the formation of structure in the early universe.
However, as discussed below, the various rate coefficients adopted by
the astrophysics community for this process differ from one another by
orders of magnitude.  This translates into orders of magnitude
uncertainties in the predicted relative abundance of H$_2$ at key
epochs. Furthermore, this uncertainty directly affects the predicted
properties of collapsing gas clouds during primordial galaxy and first
star formation.

To help address the need for accurate rate coefficients for
reaction~(\ref{eq:CTH2}), \citet{Krs02a} has recently carried out the
first quantum mechanical calculations for this process at collision
energies relevant for the cosmological formation of structure.  His
results are appropriate for current models of early universe
chemistry, which treat all molecules as being in their ground
rovibrational state.  These models have not yet evolved to the point
where they take into account the electronic and rovibrational
distribution of molecules \citep[e.g.,][]{Oh02a,Lep02a}.  Throughout
the rest of this paper only ground electronic configurations are
considered.

In \S~\ref{sec:Review}, we give a brief review of the previous
experimental and theoretical work for reaction~(\ref{eq:CTH2}).  A short
description of the theoretical calculations by \citet{Krs02a} is given in
\S~\ref{sec:Theoretical}.  In \S~\ref{sec:Results} we present our new rate
coefficient, provide a simple fitting formula that can readily be
incorporated into early universe chemistry models, and comment on our
results.  Lastly, in \S~\ref{sec:Cosmological} we briefly discuss some of
the cosmological implications of our results.

\section{Review of Previous Work}
\label{sec:Review}

\begin{figure}
\plotone{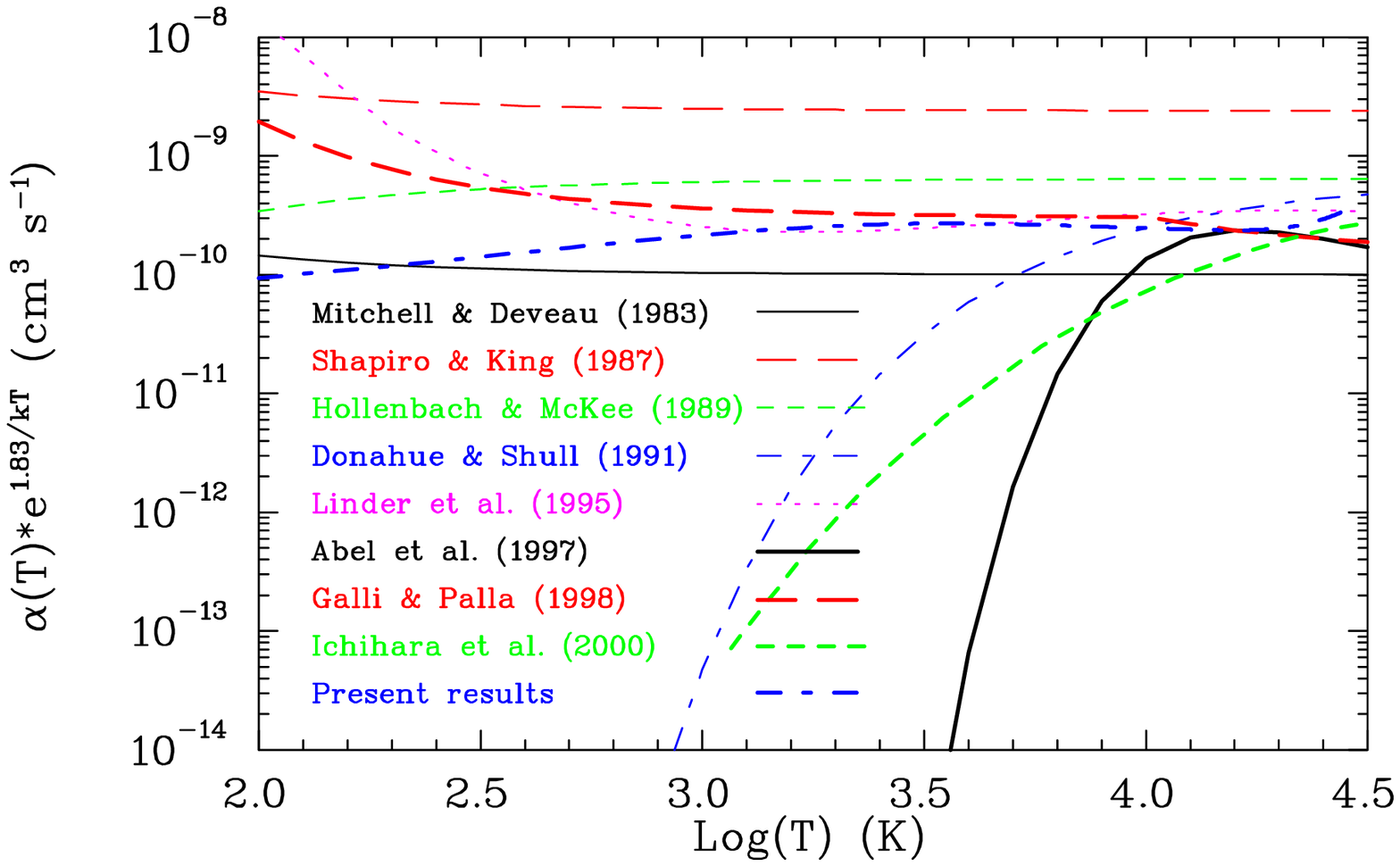}
\caption{Recommended rate coefficients for ${\rm H}_2 + {\rm H}^+ \to
{\rm H}_2^+ + {\rm H}$.  All data have been multiplied by
$\exp(1.83/kT)$ to remove the effects of the 1.83~eV threshold for
this process.}
\label{fig:h2hplusct}
\end{figure}

CT of H$^+$ on H$_2$($\nu=0,\ J=0$) is an endothermic process with a
threshold of 1.83~eV \citep{Jan87a}.  Cosmologically,
reaction~(\ref{eq:CTH2}) is important at temperatures of $kT < 1$~eV, where
$k$ is the Boltzmann constant.  Hence, it is the behavior of the cross
section just above threshold that is most important for primordial
chemistry.  The only cross section measurements in this energy range have
been carried out by Holliday, Muckerman, \& Friedman (1971).  A rate
coefficient measurement for the reverse of reaction~(\ref{eq:CTH2}) has been
carried out at a temperature of $T=300$~K by Karpas, Anicich, \& Huntress
(1979).  Rate coefficients for reaction~(\ref{eq:CTH2}) have been presented
by a number of different groups. In Figure~\ref{fig:h2hplusct} we present
the various published data multiplied by $\exp(1.83/kT)$ to remove the
effects due to the threshold for this reaction.

\citet{Mit83a}, \citet{Sha87a}, and \citet{Hol89a} appear to present
data derived using detailed balance between the forward and reverse
directions of reaction~(\ref{eq:CTH2}) and the rate coefficient
measurement for the reverse direction by \citet{Kar79a}.  However,
reading their papers and their cited sources, how their data are
actually derived is unclear. \citet{Don91a} state explicitly that
their data are derived from detailed balance between the forward and
reverse directions of reaction~(\ref{eq:CTH2}) and the measurements of
Karpas et al.  Yet, as shown in Figure~\ref{fig:h2hplusct}, these four
recommended rate coefficients all differ from one another dramatically
in both magnitude and temperature dependence, despite their all
apparently being based on detailed balanced between the forward and
reverse directions of reaction~(\ref{eq:CTH2}) and despite the use of
the same single temperature laboratory measurement.

We note that the application of detailed balance is only strictly
valid for state-to-state reactions, i.e., when the $v,J$ level of the
reactant {\it and} product molecules are known.  While typical
laboratory conditions are such that $v=0$ and $J$ is likely to be
small for the reactant, the product $v,J$ is usually unknown. For
example, Krsti\'c (2002) finds that for the reverse of reaction (1), the
product H$_2$ is primarily formed into $v=4$, not $v=0$. Therefore,
estimation of reaction~(\ref{eq:CTH2}) by the application of detailed
balance to the measured rate coefficient for the reverse reaction
gives the rate coefficient for H$_2$($v=4$), which can be as much as
an order of magnitude larger than for H$_2$($v=0$).  Rate coefficients
which are estimated by detailed balance are therefore suspect.

Some groups have derived the needed rate coefficients using the cross
section measurements of \citet{Hol71a}.  \citet{Abe97a} integrated the
recommended cross section data of \citet{Jan87a}, which are based on the
measurements by Holliday et al.  \citet{Gal98a} directly integrated the
data of Holliday et al.  Hence, it is unclear why the rate coefficients of
Abel et al.\ and Galli \& Palla differ so dramatically in both temperature
dependence and absolute magnitude.

Linder, Janev, \& Botero (1995) have presented a recommended cross section
for reaction~(\ref{eq:CTH2}) that we have integrated to produce a rate
coefficient using the method discussed in \S~\ref{sec:Theoretical}.  
Their cross section is based primarily on published experimental work for
both this reaction and for the isotopically identical reaction
\begin{equation}
\label{eq:CTD2}
{\rm D}^+ + {\rm D}_2(\nu=0,J=0)  \to {\rm D}(1s) + {\rm D_2}^+.
\end{equation}
But as pointed out by \citet{Krs02a}, several eV above the threshold, the
recommended cross section of Linder et al.\ is over an order of
magnitude larger than the measurements of \citet{Hol71a}.  So the
relative agreement between the rate coefficients of \citet{Lin95a} and
\citet{Gal98a}, despite their being based on cross section data sets
that differ significantly from one another, is surprising.

More recently, theoretical calculations have been carried out using the
classical trajectory-surface-hopping (TSH) model of Ichihara, Iwamoto, \&
Janev (2000).  In Figure~\ref{fig:h2hplusct} we have plotted their results as
a function of plasma temperature for an H$_2$ temperature of 0.1~eV.  We
expect that at this H$_2$ temperature their results are nearly equivalent
to the results for ground state H$_2$.  The TSH method, however, is
expected to provide only qualitative results for center-of-mass (CM)
energies $E_{CM} \lesssim 10$~eV.  Early universe chemical models need
rate coefficients for these reactions at temperatures $\lesssim 10^4$~K.  
This corresponds to $E_{CM} \lesssim 1$~eV.

To summarize, it is clear that more sophisticated theoretical calculations
and new laboratory measurements for reaction~(\ref{eq:CTH2}) are needed.

\section{Theoretical Method}
\label{sec:Theoretical}

Cross sections for vibrationally resolved charge-transfer of protons
with H$_{2}(\nu _{i})$ and of atomic hydrogen with H$_{2}^{+}(\nu
_{i})$ have been obtained by solving the Schr\"{o}dinger equation for
the nuclear and electronic motions on the two lowest diabatic
electronic surfaces of H$_{3}^{+}$ \citep{Krs02a}.  The calculations
were performed using a fully-quantal, coupled-channel approach by
expanding the nuclear wave functions in a large vibrational basis of
all discrete H$_{2}(\nu=0-14)$ and H$_{2}^{+}(\nu^{\prime }=0-18)$
states and corresponding discretized continua pseudostates (altogether
900 states of positive and negative energy). As a consequence, all
inelastic processes (CT, excitation, and dissociation)
were calculated on the same footing, enabling both proper population
dynamics and normalization of the full $S$-matrix
\citep{Krs02a,krstic03a}.  Discretized vibrational continua and large
configuration spaces (40 atomic units [a.u.] in length) were used
along with the bound states to account for transitions through the
``closed'' channels and for nuclear particle exchange.  The price paid
of using a diabatic vibrational basis is the large number of closed
channels needed to achieve the convergence of the cross sections
(typically 200 at lower energy to 600 at the highest energies
considered here).  This is partially a consequence of the large number
of quasi-continuum states for representing particle-exchange channels,
which are certainly present in the large, $40~{\rm a.u.} \times 
40~{\rm a.u.}$, quantization box.

The main approximation used was the sudden approximation for rotations,
often referred as the Infinite Order Sudden Approximation (IOSA).  This
technique freezes target molecule rotations during the collision and
then post-collisionally averages the cross sections over all possible
molecular orientations
\citep{pack74,secr,chu,kupper76,schatz76,khar78,kour79,baer85,baer87,sid}.
This approximation set the lower limit of the appropriate range for
calculations to a fraction of eV.  In principle, there is no upper limit.  
With the IOSA prescriptions and by expansion of the nuclear functions in
the vibrational basis, the resulting system of coupled second-order
ordinary differential equations in $R$ (the distance from the projectile
to the molecular CM)  reduces to uncoupled equations for each partial wave
$\ell$ of the projectile CM motion. These were solved using as many
partial waves as needed until convergence of the cross section was
achieved (going up to $\ell=600$ for the highest value of $E$). The proper
plane-wave boundary conditions were applied at entrance-exit of the
reactant configuration (i.e., at $R=R_{\max }=40$~a.u.), utilizing
multichannel logarithmic derivatives, for both open and closed channels
\citep{john}. The open-open submatrix of the resulting $K$-matrix is then
used to obtain the $S$-matrix for each $\ell$.

A number of different definitions of the plasma rate coefficient exist
in the literature.  Here we have calculated the rate coefficients
$\alpha(T)$ taking the Maxwellian average of the cross section
$\sigma(E)$ over the relative energy distribution of the H$^+$ and
H$_2$ particles using
\begin{equation}
\alpha(T) = \frac{1}{\sqrt{\pi \mu}} \Bigl(\frac{2}{kT}\Bigr)^{3/2} 
\int_{E_{t}}^{\infty} \sigma(E) E \exp(-E/kT) dE
\end{equation}
where $\mu$ is the reduced mass of the system, $E_{t}$ is the
threshold energy for the considered process, and $E=\mu v^2/2$ is the
CM energy. Our calculated data for $\sigma$ extend from 2 to 9.5
eV.  We integrated from 1.83 to 9.5 eV, by assuming $\sigma
(E_{t})=0$.  A spline-fit parabolic curve was used to interpolate
$\sigma$ between the calculated values.  We estimate that the cutoff
of the integration at 9.5 eV introduces less than 1\% error in our
results.

Further details of this calculation can be found elsewhere
\citep{Krs02a}.  Also available for both the H$^{+}$+H$_{2}$ and
H+H$_{2}^{+}$ collision systems are the final and initial vibrational
state resolved cross sections for CT \citep{Krs02a},
excitation \citep{Krs02a}, dissociation \citep{krstic03a}, and elastic
transport \citep{krstic99,krstic03b}. The three-body association rate
coefficients for H$^{+}$+H+H forming H$_2$ in the temperature range
200-20,000 K have also been calculated \citep{krstic03c}. All the
cross section data can be downloaded in tabular and graphic form from
www-cfadc.phy.ornl.gov.

\section{Results and Discussion}
\label{sec:Results}

We have calculated the rate coefficient for reaction~(\ref{eq:CTH2}) for
temperatures of $10^2-3\times 10^4$~K.  This covers the range of
temperatures for which reaction~(\ref{eq:CTH2}) is cosmologically
important.  We present our results in Figure~\ref{fig:h2hplusct}.

The calculated CT rate coefficient has been fitted using
\begin{equation}
\alpha(T) = \exp(-a/T)\sum_{i=0}^{7} b_i (\ln T)^i.
\label{eq:fit}
\end{equation}
The best fit values are listed in Table~\ref{tab:fitparameters}.  The
fit is accurate to better than 4\% for $10^2 \le T \le 3\times 10^4$~K.

Our calculated results represent the most sophisticated theoretical
treatment of reaction~(\ref{eq:CTH2}) at the present time.  Previous
rate coefficient calculations, based on detailed balance or the TSH
method, are unreliable for the reasons discussed in
\S~\ref{sec:Review} and cannot be used to verify the validity of our
results.  Similarly, \citet{Lin95a} raise questions about the
reliability of the laboratory measurements by \citet{Hol71a}.  Hence,
in our opinion, rate coefficients derived from the cross section data
of either \citet{Hol71a} or \citet{Lin95a} cannot be used to verify
the reliability of our calculated results.  In short, new laboratory
measurements are needed to benchmark these new calculations.

\section{Cosmological Implications}
\label{sec:Cosmological}

Using the models of \citet{Oh02a}, we have investigated how our new results
for reaction~(\ref{eq:CTH2}) affect the predicted gas temperature and H$_2$
relative abundance of a collapsing gas cloud during the epoch of
primordial galaxy and first star formation.  Calculations have been
carried out first in the absence of any external UV field and then assuming
an external field of $10^{-21}$ ergs
s$^{-1}$ cm$^{-2}$ sr$^{-1}$ in the Lyman-Werner bands of H$_2$ (11.2-13.6
eV).  The flux chosen for the UV field is roughly that needed to
reionize the universe (Haiman, Abel, \& Rees 2000). When using an external
UV field, we assumed a hydrogen column density of 10$^{22}$~cm$^{-2}$
to mimic the interior of a dense, initially neutral clump.  We used a
cosmic H/He ratio of $Y_{He}=0.24$.  For reaction~(\ref{eq:CTH2}), we have
used the adopted data of \citet{Sha87a}, \citet{Abe97a}, and our new
results.  The rest of the chemical network is from \citet{Oh02a}.

\begin{figure}
\plotone{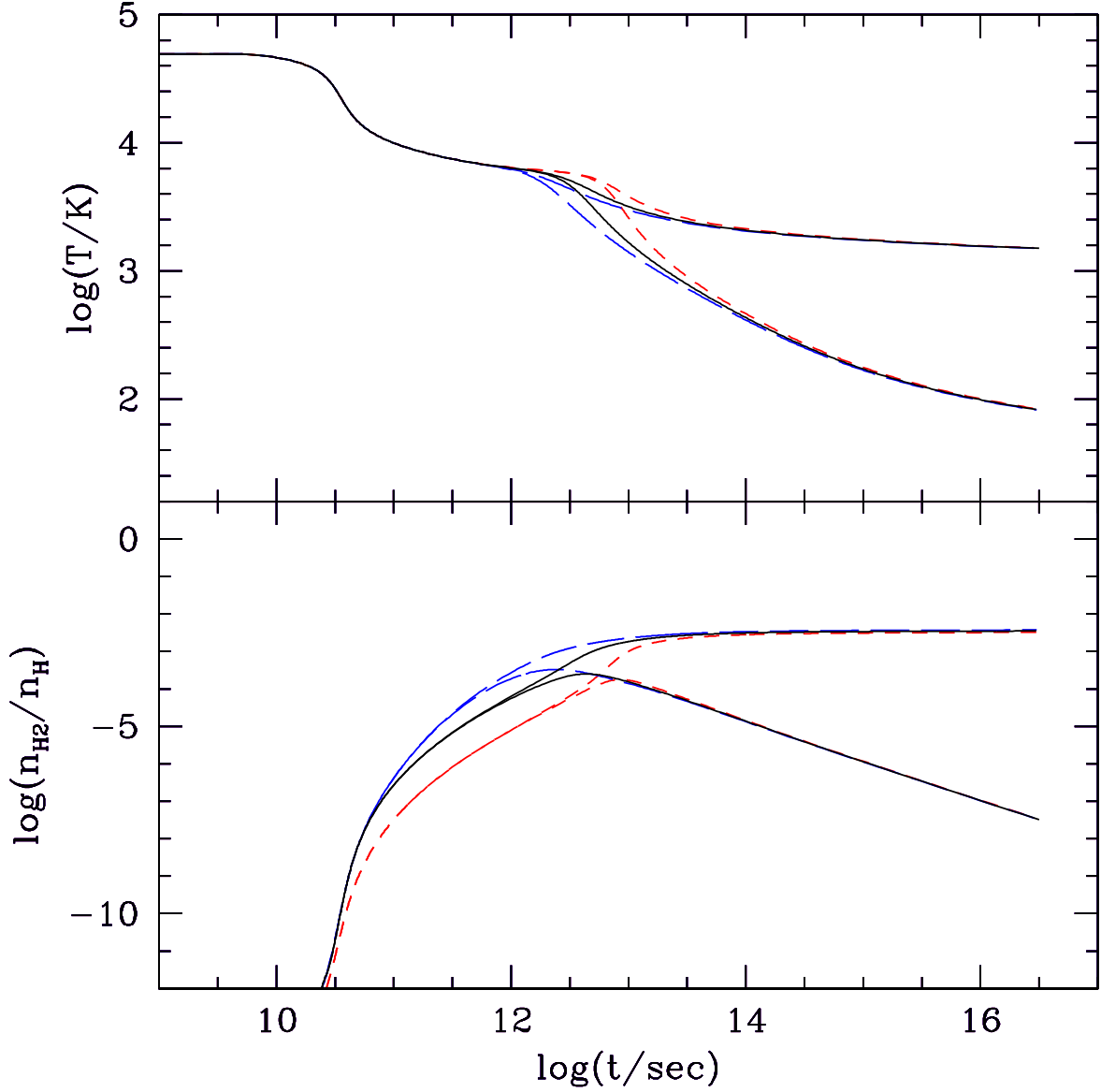}
\vfill
\eject
\caption{Calculated gas temperature ({\it top}) and H$_2$ relative abundance
({\it bottom}) for a parcel of gas starting with a density of 10 cm$^{-3}$.
In the top(bottom) plot, the lower(upper) set of curves correspond to
no external UV field and the upper(lower) set of 3 curves to a field
of $10^{-21}$ ergs s$^{-1}$ cm$^{-2}$ sr$^{-1}$ in the Lyman-Werner
bands of H$_2$ (11.2-13.6 eV).  The short dashed curves use the
rate coefficient for reaction~(\ref{eq:CTH2}) of Shapiro \& Kang (1987),
the long dashed curves use the values of Abel et al.\ (1997),
and the solid curves uses the present results.  See 
\S~\ref{sec:Cosmological} for further details.}
\label{fig:zoltan}
\end{figure}

In Figure~\ref{fig:zoltan}, we show the predicted gas temperature and
H$_2$ relative abundance using these three different rate
coefficients.  We note that using the rate coefficient of
\citet{Gal98a} yields results similar to those using our new data.
Here our model follows a parcel of gas at an initial density of 10
cm$^{-3}$.  This is about $10^4$ times denser than the background
intergalactic medium at a redshift $z \approx 20$.  Assuming that $z
\approx 20$ at $t=0$~s, then the evolution of the gas is followed
until $t=10^{16}$~s. Using the {\it WMAP} results and the LCDM cosmological
model of \citet{Spe03a}, this corresponds to a final redshift of $z
\approx 5$.  The asymptotic values for the predicted gas temperature
and H$_2$ relative abundance are little affected by the atomic physics
uncertainties \citep[a well known result,
e.g.,][]{Abe97a,Gal98a,Oh02a}.  But as can be readily seen from the
figure, for times between $10^{11}$ and $10^{13}$~s, both the
predicted gas temperature and H$_2$ relative abundance are highly
sensitive to the rate coefficient chosen.  These timescales are short
compared to cosmic timescales but could well be relevant for the
formation of the first stars \citep{Hai00a}.  For example $10^{13}$~s
corresponds to the dynamical time for gas at over-densities of a few
times $10^5$ relative to the cosmic mean at redshift $z \approx 20$,
and the actual star formation process may proceed on even shorter
time-scales \citep{Abe02a}.  Our new theoretical results for
reaction~(\ref{eq:CTH2}) significantly reduce the uncertainties in the
predicted gas temperature and H$_2$ relative abundance due to the
adopted rate coefficient.

\acknowledgements

D.W.S was supported in part by the NASA Space Astrophysics Research and
Analysis Program grant NAG5-5420 and the NSF Galactic Astronomy Program
grant AST-0307203.  P.S.K was supported by the US Department of Energy,
Office of Fusion Energy Sciences, through Oak Ridge National Laboratory,
managed by UT-Battelle, LLC under contract DE-AC05-00OR22725.  Z.H. was
supported in part by the NSF Extragalactic Astronomy \& Cosmology Program
grant AST-0307291.  P.C.S. acknowledges support from the NSF Extragalactic
Astronomy \& Cosmology Program grant AST-0087172.

\vfill
\eject

\vfill
\eject

\vfill
\eject

\begin{deluxetable}{cc}
\tablecaption{Fit parameters for Equation~\ref{eq:fit}.  The units for $a$ 
are K and for $b_i$ they are cm$^3$~s$^{-1}$.
\label{tab:fitparameters}}
\tablewidth{0pt}
\tablehead{
\colhead{Parameter} & \colhead{Value}}
\startdata
$a$      &  2.1237150E+04 \\
$b_0$    & -3.3232183E-07 \\
$b_1$    &  3.3735382E-07 \\
$b_2$    & -1.4491368E-07 \\
$b_3$    &  3.4172805E-08 \\
$b_4$    & -4.7813720E-09 \\
$b_5$    &  3.9731542E-10 \\
$b_6$    & -1.8171411E-11 \\
$b_7$    &  3.5311932E-13 \\
\enddata
\end{deluxetable}

\end{document}